\documentclass[twocolumn,aps,prd,showpacs]{revtex4}
\usepackage[T1]{fontenc}
\usepackage[latin9]{inputenc}
\usepackage{bm}
\usepackage{amsmath}
\usepackage{amssymb}
\usepackage{graphicx}
\usepackage{esint}
\usepackage{hyperref}

\makeatletter

\@ifundefined{textcolor}{}
{%
 \definecolor{BLACK}{gray}{0}
 \definecolor{WHITE}{gray}{1}
 \definecolor{RED}{rgb}{1,0,0}
 \definecolor{GREEN}{rgb}{0,1,0}
 \definecolor{BLUE}{rgb}{0,0,1}
 \definecolor{CYAN}{cmyk}{1,0,0,0}
 \definecolor{MAGENTA}{cmyk}{0,1,0,0}
 \definecolor{YELLOW}{cmyk}{0,0,1,0}
}

\makeatother

\begin{document}

\title{Rederivation of the Casimir force under the completeness relation of continuum operator}

\author{Gao Xianlong$^{1}$ and Kelin Wang$^{2}$}

\affiliation{$^{1}$Department of Physics, Zhejiang Normal University, Jinhua
321004, China}
\affiliation{$^{2}$  Department of Modern Physics, University of Science and Technology of China,
Hefei 230026, China}

\date{\today}
\begin{abstract}
Casimir effects manifests that, the two closely paralleled plates, generally produce a macroscopic attractive force due to the quantum vacuum fluctuations of the electromagnetic fields. The derivation of the force requires an {\it artificial} regulator by removing the divergent summation. By including naturally a spectrum density factor, based on the observation that an incomplete eigenvectors of observable, such as the eigenstates for the photons in the free field, can form a complete set of eigenvectors by introducing a unique spectrum transformation, an alternative way is presented to rederive the force,
without using a regulator. As a result, the Casimir forces are obtained with the first term $-\pi^2 \hbar c/(240 a^4)$ attractive, and the second one, $-\pi^4 \hbar c^3 \sigma^2/(1008 a^6)$, also attractive but smaller, with $a$ the plate separation, and $\sigma$ a to-be-determined small constant number in the spectrum density factor.

\end{abstract}


\maketitle

\section{Introduction}
The well known Casimir effects~\cite{Casimir,Gambassi-review}, demonstrates that the quantum vacuum fluctuation of the electromagnetic fields produces a macroscopic attractive force between the two closely paralleled plates. Or, more generally, the zero point energy of the confined photon fields contributes to the force, which is $-\pi^2\hbar c/(240 a^4)$ with $a$ being the plate separation, arising when the virtual particles are excluded
from the space between the plates as the separation $a$ is smaller than the wavelength of the particles. The Casimir forces is precisely measured by Lamoreaux in 1997 between parallel conducting plates to about 1\% precision~\cite{Lamoreaux}.

Since that, people realized that quantum vacuum fluctuations had measurable consequences. For example, the atomic Casimir effect can account for the Lamb shift of spectra~\cite{Billaud} and modify the magnetic moment for the electron~\cite{Narozhny}. This type of vacuum fluctuations is now vital to our understanding of nature.

Recently, renewed attention has been focused on the Casimir effects branching out in various fields ranging from nanoscopic physics~\cite{Bordag-review} to cold atomic physics~\cite{coldatom}.
The Casimir pressure between two gold-coated plates was measured with an error of 0.2\% at $d =160nm$~\cite{Decca-gold} and more elegant measurement was done between a metalized sphere and flat plate with the plate-sphere surface separations from 0.1 to 0.9 $\mu m$~\cite{Mohideen}.

Further studies concern the gravitational Casimir effect with nonidealized boundary conditions~\cite{Gravitational-CE} and the fermion Casimir effect triggered due to the statistics obeyed by fermions as opposed to bosons~\cite{Flachi}.
When the plates moves at relativistic speeds, real photons are produced while the speed of the pairs of the virtual particles does not match the speed of the plates, termed as the dynamic Casimir effect, recently observed in a superconducting circuit~\cite{{dynamic CE}}.
Ultracold atomic gases provide one of the most versatile platforms
for realizing exotic many-body quantum states of matter, due to their unprecedented tunability and controllability in almost all aspects of the system parameters\cite{Bloch2008,Giorgini2008}. Recently, Casimir effects are studied in a cold atomic sample
of dilute Rydberg atoms trapped in front of a rough substrate~\cite{coldatom}.

Originally derived by using the quantum-mechanical perturbation theory to fourth order in $e$~\cite{Casimirbook}, the Casimir force, in the standard approach, is calculated by computing the change in the zero-point
energy $E$ per unit area of the electromagnetic field when the separation between perfectly paralleled conducting plates is changed, that is, $F_c=-\partial E/\partial a $. This derivation is mathematically much simpler.
However, debate still exists. Schwinger had pointed out that the Casimir effects can be explained without reference to zero-point energies or even
to the vacuum~\cite{Schwinger}. Also Jaffe realized that the concept of zero point fluctuations is not a necessity but a heuristic and calculational aid in the description of the Casimir effect~\cite{Jaffe}.

Moreover, in the calculation of the vacuum energy, infinite sums over the momenta are taken which lead to divergence, and therefore, artificial regulators are needed, to remove the divergence. For example, the zeta-function, heat kernel, and Gaussian regulator are introduced. In this paper, partially motivated by the rapid experimental progress in the Casimir effect, and aiming at avoiding the regulations of the infrared divergences in the zero energy, we rederive the Casimir force by considering the completeness relation of continuum spectrum of photons, where a spectrum density factor is naturally included to count for the number of eigenstates for the different momentum of the photons.

The rest of the paper is organized as follows: in Sec. II, we introduce a spectrum density of free photons in the free field or a statistic weight factor for photons of a momentum. In Sec. III, we use this spectrum density to rederive the Casimir effect. In addition, we obtain a small correction to the Casimir force, which in turn settles down the undetermined constant in the spectrum density. 
Section IV is devoted to conclusions and outlooks.

%

\section{Spectrum density of free photons in the free field}
We consider the typical Casimir effect with a pair of uncharged conducting metal plates at distance $a$ apart.
Assuming the parallel plates lie in the $xy$-plane, then, the virtual photons which constitute the vacuum field of quantum electrodynamics are free in $xy$ and confined in $z$ directions. The standing waves between the metal plates are,
\begin{equation}
\psi_n(x,y,z;t) = e^{-i\omega_{k,n}t} e^{ik_xx+ik_yy} \sin \left(k_n z \right)~,
\end{equation}
with $k_x, k_y\in (-\infty, \infty)$ the wave vectors free in the $xy$-direction. However, the boundary condition in the $z$-direction $\psi_n(x,y,0;t)=\psi_n(x,y,a;t)$ requires,
\[
k_n=\frac{n\pi}{a},~(n=1,2,...).
\]
The frequency of the wave is
\begin{equation}
\omega_{k,n} = c\sqrt{k^2 + \frac{n^2\pi^2}{a^2}}~,
\label{spectrum}
\end{equation}
with $k^2={k_x}^2 + {k_y}^2$.
The vacuum energy (zero energy) per area is then,
\begin{equation}
E(a) = 2 \int\int \frac{ d^2k}{(2\pi)^2} \sum_{n=1}^\infty \frac{\hbar}{2} \omega_{k,n}~,
\label{Eofa}
\end{equation}
where a factor of $2$ is responsible for the two possible polarizations of the wave, and from which the Casimir force can be calculated,
\begin{equation}
F_c=\frac{\partial E(a)}{\partial a}~.
\end{equation}
By noticing the summation over $n$ in Eq. (\ref{Eofa}), the result is {\it clearly infinite}!
The computation of the Casimir force also leads to infinite sums, and therefore, requires regularization.
The divergent sum for vacuum energy can be decomposed into an infinite and a finite part.
Usually a regulator, such as, of zeta-function, heat kernel, or Gaussian, is introduced to make the expression finite, and in the end it will be removed. Normally the finite part doesn't depend on the choice of the regulator.
Actually, Jared Kaplan has proofed in this course note~\cite{Kaplan} that, any regulator $f(x)$ in $E(a)$ gives
\begin{equation}
E(a) = \hbar\int \int\frac{ d^2k}{(2\pi)^2} \sum_{n=1}^\infty f(\frac{n}{L\Lambda}) \omega_{k,n}
\sim -\frac{\hbar c\pi^2 f(0)}{720 a^3}~,
\label{Eofa1}
\end{equation}
and thus the correct attractive Casimir force
as long as $ f(0) = 1 $. Besides, another two requirements on $f(x)$ are that,
the ultra high energy, short distance modes are irrelevant for the physics and the short distance regulator function $f(x)$ does not change the modes at very long distances, where the Casimir effect actually arises. Here, $L\gg a$ and $\Lambda$ is the high momentum cutoff.



However, to derive the Casimir force, it is not a necessity to induce the regularization.
For example, the UV cut-off is possible~\cite{Ichinose}. In Casimir's original paper~\cite{Casimir}, he
compared the situation in which the plate is at a small distance and the
situation in which it is at a very large distance. The difference between the two gives an finite attractive force.

In this paper, we present an alternative way to rederive the force by including a spectrum density factor
without introducing a regulator (discussed in greater detail below and in the Appendix).
Before going into the results, we have to discuss the completeness relation of the photon in the free space~\cite{Hawton}.

For the photon in the free field, the eigenvector set of its momentum and energy is, $\{\vert {\bf k}, \lambda\rangle^{(1)}  \}$,
where ${\bf k}$ is its wave vector (or momentum), and $\lambda$ its two transverse polarizations.
The eigenvector usually takes,
\begin{equation}
\vert {\bf k}, \lambda\rangle^{(1)} \sim e^{-i{\bf k}\cdot {\bf x}}~.
\label{Eq:photon}
\end{equation}

Same as the two examples in the Appendix, the set of eigenstates of the continuum spectra is not unique, and usually, incomplete.
This eigenstate can not be used as a basis, and as a result, the physical quantities based on it will lead to divergence.
From the Appendix, we know that an incomplete eigenvectors of observable can be transformed into a unique complete set by introducing a spectrum density factor. Thus, we change Eq. (\ref{Eq:photon}) into,
\begin{eqnarray}
\vert {\bf k}, \lambda\rangle &=& f({\bf k},\lambda) \vert {\bf k}, \lambda\rangle^{(1)}\nonumber\\
&=& f({ k}) \vert {\bf k}, \lambda\rangle^{(1)}~,
\label{Eq:photon2}
\end{eqnarray}
where $f({\bf k},\lambda)$ is the spectrum density factor. The second equality is due to the isotropic of the space.
After the transformation, the set of eigenstates for photons in free space is given by Eq. (\ref{Eq:photon2}) satisfying the completeness relation, and
$f({ k})$ is the spectrum density or the statistic weight factor for photons of momentum $k$, to be determined by the relevant experiment, for instance, the experiments measuring the Casimir pressure between two gold-coated plates~\cite{Decca-gold}.
From the above analysis, we conclude that the number of eigenstates within ${\bf k}\rightarrow {\bf k}+d{\bf k}$ is $f({ k}) \vert {\bf k}, \lambda\rangle^{(1)}d{\bf k}$.

\section{Casimir effect}
In this section, we will rederive the Casimir effect by considering the spectrum density factor for photons within momentum ${\bf k}+d{\bf k}$. For the definite $n$, Eq. (\ref{spectrum}) gives,
\begin{equation}
kdk=\frac{1}{c^2}\omega_{k,n} d\omega_{k,n}~,
\label{spectrum2}
\end{equation}
and Eq. (\ref{Eofa}) is rewritten as,
\begin{eqnarray}
E(a) &= \hbar &\sum_{n=1}^\infty \int\int \frac{ k dk d\varphi}{(2\pi)^2} \omega_{k,n}~,
\nonumber\\
&= & \frac{\hbar }{2\pi c^2} \sum_{n=1}^\infty \int_{cn\pi/a}^{\infty} (\omega_{k,n})^2 d\omega_{k,n}~.
\label{Eofa2}
\end{eqnarray}
As we discussed in Sec. II, a spectrum density factor $f(\omega_{k,n})$ is missing in Eq. (\ref{spectrum2}) and thus the above
equation has to be,
\begin{eqnarray}
E(a) =  \frac{\hbar }{2\pi c^2} \sum_{n=1}^\infty \int_{cn\pi/a}^{\infty}f(\omega_{k,n}) (\omega_{k,n})^2 d\omega_{k,n}~.
\label{Eofa3}
\end{eqnarray}
Now we derive the analytic expression of $E(a)$ by assuming $f(\omega)=e^{-\sigma  \omega}$, where $\sigma$ has the unit of time, and is a constant to be determined by the experiments. We want to emphasize that, the scheme we proposed here, is obviously different from the normally adopted regulation procedure, where the infinitesimal number is chosen to be $\sigma\rightarrow 0$ at the end of the calculation. 

By doing the integration over $\omega$, we achieve,
\begin{eqnarray}
E(a) &=&  \frac{\hbar }{2\pi c^2} \frac{d^2}{d\sigma^2} \sum_{n=1}^\infty \int_{cn\pi/a}^{\infty}e^{-\sigma\omega_{k,n}} d\omega_{k,n}
\nonumber\\
&=&\frac{\hbar }{2\pi c^2} \frac{d^2}{d\sigma^2} \sum_{n=1}^\infty \frac{1}{\sigma}e^{-\frac{cn\pi \sigma}{a}}
\nonumber\\
&=&\frac{\hbar }{2\pi c^2} \frac{d^2}{d\sigma^2} \frac{1}{\sigma} \left[\frac{1}{1-e^{-c\pi\sigma/a}}-1\right]
\label{Eofa4}
\end{eqnarray}

Making use of the expansion,
\begin{equation}
\frac{1}{1-e^x}=-\sum_{n=0}^\infty B_n \frac{x^{n-1}}{n!}~,~|x|\in (0, 2\pi)~,
\label{Bernoulli}
\end{equation}
where $B_n$ is the Bernoulli numbers~\cite{RG}, we obtain the energy,
\begin{eqnarray}
E(a) &=& 3B_0 \frac{\hbar a}{\pi^2 c^3\sigma^4}-(1+B_1)\frac{\hbar}{\pi c^2\sigma^3}
\nonumber \\
&&+B_4\frac{\pi^2\hbar c}{24 a^3}-B_5 \frac{\pi^3 \hbar c^2\sigma}{40 a^4}+B_6\frac{\pi^4\hbar c^3\sigma^2}{120 a^5}+...~.
\label{Eofa5}
\end{eqnarray}
Here, $B_0=1, B_1=-\frac{1}{2}, B_2=\frac{1}{6}, B_3=0, B_4=\frac{1}{30}, B_5=0$, and $B_6=-\frac{1}{42}$.

When the distance between the plates goes to infinity $a\rightarrow \infty$, it is the energy of the real vacuum,
\[
{\rm lim}_{a\rightarrow \infty} E(a)=  3B_0 \frac{\hbar a}{\pi^2 c^3\sigma^4}-(1+B_1)\frac{\hbar}{\pi c^2\sigma^3}~.
\]
When choosing the zero energy to be the real vacuum, we obtain the effective energy per area between the plates,
\begin{eqnarray}
E_{\rm eff}(a)=B_4\frac{\pi^2\hbar c}{24 a^3}-B_5 \frac{\pi^3 \hbar c^2\sigma}{40 a^4}+B_6\frac{\pi^4\hbar c^3\sigma^2}{120 a^5}+...~,
\label{Eofa6}
\end{eqnarray}
and the force per area,
\begin{eqnarray}
F_c&=&-\frac{\partial}{\partial a }E(a)\nonumber\\
&=&-\frac{\pi^2\hbar c}{240 a^4} -\frac{\pi^4 \hbar c^3\sigma^2}{1008 a^6}+...~.
\end{eqnarray}
It is seen that the plates do affect the virtual photons which constitute the field, and generate a net attractive force.
The first term gives the usual attractive Casimir force. The second term is a small correction to the force, which means $\sigma\sim a/c$. 
and in turn settles down the undetermined constant $\sigma$ in the spectrum density.  $\sigma$ can be determined by the precise measurement on the force in the experiment.
In all, $\sigma$ is a small constant number, which is consistent with our observation in that, only for photons with very high frequency, we can find the different behaviors through the spectrum density factor $f(\omega_{k,n})\sim 1-\sigma\omega_{k,n}$ we adopted, while in the normal circumstances, the photons, contribute almost equally even with different frequencies since the spectrum density factor scales$f(\omega_{k,n})\sim 1$.

We hope in the near experiments, 

\section{Conclusions}
We have rederived the Casimir force between the two paralleled metallic plates
by using the completeness of the eigenvector set for the photon in the free field.
We have found that without introducing a regulator or the UV-cutoff, we obtained
the Casimir force while including a spectrum density factor. The first term gives
the correct attractive Casimir force, $-\pi^2\hbar c/(240 a^4)$, and the second term,
$-\pi^4 \hbar c^3 \sigma^2/(1008 a^6)$, is also attractive and gives a small correction to the force.
The constant $\sigma$ in the spectrum density could be determined by the precise experiments
measuring the Casimir pressure between two gold-coated plates~\cite{Decca-gold} or in a cold atomic sample
of dilute Rydberg atoms trapped in front of a rough substrate~\cite{coldatom}.

In the near future, it is of interest to consider Lamb energy shift~\cite{Billaud}.
Other physics relying on the regulator or UV cutoff could also be addressed by
including the present spectrum density factor~\cite{Gravitational-CE,Flachi}.

\begin{acknowledgments}
This work was supported by the NSF of China (Grant No. 11374266)
and the Program for New Century Excellent Talents in University. 
\end{acknowledgments}

\appendix

\section{The completeness of the continuum operators}
In this Appendix, we explain that the completeness of the continuum operators could be achieved by a transformation.
We illustrate the idea by first discussing it for a discrete number operator $\hat{n}=\hat{a}^\dagger\hat{a}$,
which is Hermitian and has the following eigenvector set $\{\vert n\rangle^{(1)} \}$,
\[
\vert n\rangle^{(1)} =(\hat{a}^\dagger)^n \vert 0\rangle.
\]
However, this set does not satisfy the completeness relation,
\begin{equation}
\sum_n \vert n\rangle^{(1)} {}^{(1)}\langle n\vert=\hat{I},
\label{eq:dis-complete}
\end{equation}
due to the fact that $\vert n\rangle^{(1)} $ is not normalized. A simple transformation
solves the problem by introducing a 'spectrum density factor' $f(n)$ (here it is the normalized constant),
\begin{equation}
\vert n\rangle=f(n)\vert n\rangle^{(1)}=f(n)(\hat{a}^\dagger)^n \vert 0\rangle
\label{eq:discrete}
\end{equation}
with $f(n)=1/\sqrt{n!}$. $\{\vert n\rangle \}$ forms a complete set of eigenstates,
\[
\sum_n \vert n\rangle\langle n\vert=\hat{I}~.
\]
Thus, for the discrete spectra of observables, it is easy to form a complete set of eigenstates by normalizing them.
However, the eigenvectors of the continuum spectra are not normalizable.
Then the question arises on how to make them a complete set of eigenstates. The key is to generalize the
above transformation $f(n)$. We clarify this point by a pair of conjugate operators $(\hat{x},\hat{p})$. By introducing
the bosonic annihilation operator $\hat{a}$ and creation operators $ \hat{a}^\dagger$,
\begin{eqnarray}
\hat{x} & = & \frac{1}{\sqrt{2}} (\hat{a}^\dagger+\hat{a}),\nonumber\\
\hat{p} & = & \frac{1}{\sqrt{2}} (\hat{a}^\dagger-\hat{a}),\nonumber
\end{eqnarray}
Here, $[\hat{a}, \hat{a}^\dagger]=1$. In the $(\hat{a}, \hat{a}^\dagger)$ space, the set of eigenstates of $(\hat{x},\hat{p})$
can be expressed in details as,
\begin{eqnarray}
\vert {x}\rangle^{(1)} & = & \exp\left[-\frac{1}{2}\hat{a}^\dagger\hat{a}^\dagger+\sqrt{2}x \hat{a}^\dagger\right]
\vert 0\rangle~,\nonumber\\
\vert {p}\rangle^{(1)} & = & \exp\left[\frac{1}{2}\hat{a}^\dagger\hat{a}^\dagger+i\sqrt{2}p \hat{a}^\dagger\right]
\vert 0\rangle~.
\label{eq:continuum}
\end{eqnarray}
Here, $x$ and $p$ are eigenvalues of $\hat{x}$ and $\hat{p}$, respectively, satisfying,
\begin{eqnarray}
\hat{x}\vert {x}\rangle^{(1)}
&=&\frac{1}{\sqrt{2}}(\hat{a}^\dagger+\hat{a})\exp\left[-\frac{1}{2}\hat{a}^\dagger\hat{a}^\dagger+\sqrt{2}x \hat{a}^\dagger\right] \vert 0\rangle\nonumber\\
&=&x\vert {x}\rangle^{(1)}~,\nonumber\\
\hat{p}\vert {p}\rangle^{(1)}
&=&\frac{i}{\sqrt{2}}(\hat{a}^\dagger-\hat{a})\exp\left[\frac{1}{2}\hat{a}^\dagger\hat{a}^\dagger+i\sqrt{2}p \hat{a}^\dagger\right] \vert 0\rangle\nonumber\\
&=&p\vert {p}\rangle^{(1)}~.\nonumber
\end{eqnarray}
Eq. (\ref{eq:continuum}) is the eigenvector set of $(\hat{x},\hat{p})$, but not the complete one. Comparing to Eq. (\ref{eq:discrete}), similar transformations can be used by introducing the 'spectrum density factor' $F_1(x)$ and $F_2(p)$,
\begin{eqnarray}
\vert {x}\rangle
&=&F_1(x) \vert {x}\rangle^{(1)},\nonumber\\
\vert {p}\rangle
&=&F_2(p)\vert {p}\rangle^{(1)}~.\nonumber
\label{Eq:spectra}
\end{eqnarray}
Here, $\vert {x}\rangle$ and $\vert {p}\rangle$ satisfy the completeness relation,
\[
\int dx \vert x\rangle\langle x\vert= \hat{I},~~\int dp \vert p\rangle\langle p\vert= \hat{I}~.
\]
In the following, we derive the detailed expression of $F_1(x)$. Making use of Eq. (\ref{eq:discrete}), we project the vector $\vert n\rangle$ into $x$ space,
\begin{eqnarray}
\langle x\vert n\rangle &=&\frac{1}{\sqrt{n!}}\int^{\infty}_{-\infty}\langle x\vert \hat{a}^{\dagger n}\vert x'\rangle \langle x' \vert 0\rangle dx' \nonumber\\
&=& \frac{1}{\sqrt{2^n n!}} \int^{\infty}_{-\infty} dx' \left( x-\frac{d}{dx}\right)^n \delta(x-x') \langle x' \vert 0\rangle \nonumber\\
&=& \frac{1}{\sqrt{\sqrt{\pi}2^n n!}}\left( x-\frac{d}{dx}\right)^n e^{-x^2/2}~.
\end{eqnarray}
Making use of the Hermite polynomial,
\[
H_n(x)=e^{x^2/2}\left( x-\frac{d}{dx}\right)^n e^{-x^2/2},
\]
we achieve,
\begin{equation}
\langle x\vert n\rangle =\frac{1}{\sqrt{\sqrt{\pi}2^n n!}}e^{-x^2/2} H_n(x)~.
\end{equation}
Combined with the complete relation Eq. (\ref{eq:dis-complete}), the Fock representation of the position operator is obtained~\cite{Wuenche},
\begin{equation}
\vert x\rangle =\pi^{-1/4} \exp\left[-\frac{x^2}{2}-\frac{1}{2}\hat{a}^\dagger\hat{a}^\dagger+\sqrt{2}x \hat{a}^\dagger\right] \vert 0\rangle~.
\end{equation}
Comparing to Eq. (\ref{Eq:spectra}), the spectrum density factor $F_x(x)$ is,
\[
F_1(x)=\pi^{-1/4} \exp\left[-\frac{x^2}{2}\right]~.
\]
Similarly,
\[
F_2(p)=\pi^{-1/4} \exp\left[-\frac{p^2}{2}\right]~.
\]
From the above example, we learn that an observable has a set of eigenstates, which is orthogonal but not necessarily complete.
An incomplete eigenvectors of observable can be transformed into a complete set by introducing a spectrum transformation, which is unique.
In the main text, we will use the same technique for the photon in the free space, and keep in mind there a similar spectrum density factor is necessary for the completeness relation of free photons.


\begin{thebibliography}{10}

\bibitem{Casimir} H. B. G. Casimir, Proc. K. Ned. Akad. Wet., Ser. B {\bf 51}, 793 (1948);
H. B. G. Casimir and D. Polder, Phys. Rev. {\bf 73}, 360 (1948).

\bibitem{Gambassi-review} A. Gambassi, The Casimir effect: from quantum to critical fluctuations,
Journal of Physics: Conference Series {\bf 161}, 012037 (2009).

\bibitem{Lamoreaux}
S.K. Lamoreaux, Demonstration of the Casimir Force in the 0.6 to 6$\mu m$ Range, Phys. Rev. Lett. {\bf 78}, 5 (1997).

\bibitem{Billaud}
B. Billaud, T.-T. Truong, Lamb shift of non-degenerate energy level systems placed between two infinite parallel conducting plates, J. Phys. A: Math. Theor. {\bf 46}, 025306 (2013).

\bibitem{Narozhny}
N. B. Narozhny, A. M. Fedotov, and Yu. E. Lozovik, Dynamical Lamb effect versus dynamical Casimir effect, Phys. Rev. A {\bf 64}, 053807 (2001);
A. K. Bhatia and Richard J. Drachman, Relativistic, retardation, and radiative corrections in Rydberg states of lithium,
Phys. Rev. A {\bf 55}, 1842 (1997).

\bibitem{Bordag-review}
M. Bordag, U. Mohideen, and V. M. Mostepanenko, Casimir effect for two spheres in a wormhole spacetime, Phys. Rep. {\bf 353}, 1 (2001).

\bibitem{coldatom}
G. A. Moreno, R. Messina, D. A. R. Dalvit, A. Lambrecht, P. A. Maia Neto, and S. Reynaud, Disorder in Quantum Vacuum: Casimir-Induced Localization of MatterWaves, Phys. Rev. Lett. {\bf 105}, 210401 (2010).

\bibitem{Decca-gold}
R. S. Decca, D. L\'{o}pez, E. Fischbach, G. L. Klimchitskaya, D. E. Krause, and V. M. Mostepanenko, Tests of new physics from precise measurements of the Casimir pressure between two gold-coated plates, Phys. Rev. D {\bf 75}, 077101 (2001).

\bibitem{Mohideen}
U. Mohideen and A. Roy, Precision Measurement of the Casimir Force from 0.1 to 0.9$\mu m$, Phys. Rev. Lett. {\bf 81}, 4549 (1998).

\bibitem{Gravitational-CE}
James Q. Quach, Gravitational Casimir Effect, Phys. Rev. Lett. {\bf 114}, 081104 (2015).

\bibitem{Flachi}
Antonino Flachi, Strongly Interacting Fermions and Phases of the Casimir Effect, Phys. Rev. Lett. {\bf 110}, 060401 (2013).

\bibitem{dynamic CE}
C.M. Wilson, G. Johansson, A. Pourkabirian, J.R. Johansson, T. Duty, F. Nori, and P. Delsing, Observation of the Dynamical Casimir Effect in a Superconducting Circuit, Nature {\bf 479}, 376 (2011).

\bibitem{Bloch2008}I. Bloch, J. Dalibard, W. Zwerger, Many-body physics with ultracold gases, Rev. Mod. Phys.
\textbf{80}, 885 (2008).

\bibitem{Giorgini2008}S. Giorgini, L. P. Pitaevskii, and S. Stringari, Theory of ultracold atomic Fermi gases,
Rev. Mod. Phys. \textbf{80}, 1215 (2008).

\bibitem{Casimirbook} M. Bordag, G. L. Klimchitskaya, U. Mohideen, and V. M. Mostepanenko, Advances in the Casimir Effect (Oxford
University Press, Oxford, 2009).

\bibitem{Schwinger}
J. Schwinger et al, Casimir effect in dielectrics, Ann., Phys. (N.Y.) {\bf 115}, 1 (1978).

\bibitem{Jaffe}
R.L. Jaffe, The Casimir Effect and the Quantum Vacuum, Phys. Rev. D {\bf 72} 021301(R) (2005).

\bibitem{Wuenche}
H-Y. Fan and J.R. Klauder, Eigenstates of two particlesÕ relative position and total momentum, Phys. Rev. A {\bf 49}, 704 (1994);
W.M. Zhang, D. H. Feng, and R. Gilmore, Coherent states: Theory and some application, Rev. Mod. Phys. {\bf 62}, 867 (1990);
A. W\"{u}nsche, About integration within ordered products in quantum optics, J. Opt. B: Quantum Semiclass. Opt. {\bf 1}, R11 (1999).

\bibitem{Ichinose}
S. Ichinose, Casimir Energy of 5D Electromagnetism and New Regularization Based on Minimal Area Principle, Prog. Theor. Phys. {\bf 121}, 727 (2009).

\bibitem{Hawton}
M. Hawton, Phys. Rev. A {\bf 59}, 954 (1999);
M. Hawton, Photon position eigenvectors lead to complete photon wave mechanics,
Proc. SPIE 6664, The Nature of Light: What Are Photons?

\bibitem{Kaplan}
J. Kaplan, QFT Lectures Notes, http://www.pha.jhu.edu/$\sim$jaredk/QFTCourseNotes.pdf.

\bibitem{RG}
 I.S. Gradsbteyn, L.M. Ryzbik, Table of integrals, series, and products, 6th edition (Singapore, 2000).

\end{thebibliography}
\end{document}